\documentclass[aps,prl,showpacs,twocolumn,preprintnumbers,superscriptaddress,nofootinbib,floatfix,10pt]{revtex4-1}
\usepackage{amsfonts,amssymb,stmaryrd,latexsym,amsmath,braket}
\usepackage{graphicx,subfigure}
\usepackage{comment}
\usepackage{times}
\usepackage{slashed}
\usepackage{bm}
\usepackage{braket}
\usepackage{chapterbib}

\newcommand{\beginsupplement}{%
        \setcounter{table}{0}
        \renewcommand{\thetable}{S\arabic{table}}%
        \setcounter{figure}{0}
        \renewcommand{\thefigure}{S\arabic{figure}}%
        \setcounter{equation}{0}
        \renewcommand{\theequation}{S\arabic{equation}}%
     }

\makeatletter
\let\saved@includegraphics\includegraphics
\AtBeginDocument{\let\includegraphics\saved@includegraphics}
\makeatother

\begin{document}

\bibliographystyle{naturemag}

\title{\textit{Ab initio} nuclear thermodynamics}

\author{Bing-Nan Lu}

\affiliation{Facility for Rare Isotope Beams and Department of Physics and
Astronomy,
Michigan State University, MI 48824, USA}

\author{Ning Li}

\affiliation{Facility for Rare Isotope Beams and Department of Physics and
Astronomy,
Michigan State University, MI 48824, USA}

\author{Serdar Elhatisari}

\affiliation{Faculty of Engineering, Karamanoglu Mehmetbey University, Karaman
70100, Turkey}

\author{Dean Lee}

\affiliation{Facility for Rare Isotope Beams and Department of Physics and
Astronomy,
Michigan State University, MI 48824, USA}

\author{Joaqu{\'i}n~E.~Drut}

\affiliation{Department of Physics and Astronomy, University of North Carolina,
Chapel Hill, North Carolina 27599-3255, USA }

\author{Timo A. L\"ahde}

\affiliation{Institute for Advanced Simulation, Institut f\"ur Kernphysik,
and
J\"ulich Center for Hadron Physics, Forschungszentrum J\"ulich,
D-52425 J\"ulich, Germany}

\author{Evgeny Epelbaum}

\affiliation{Ruhr-Universit{\"a}t Bochum, Fakult{\"a}t f{\"u}r Physik und
Astronomie,
Institut f{\"u}r Theoretische Physik II, D-44780 Bochum, Germany}

\author{Ulf-G. Mei{\ss}ner}

\affiliation{Helmholtz-Institut f\"ur Strahlen- und Kernphysik and Bethe
Center
for Theoretical Physics, Universit\"at Bonn, D-53115 Bonn, Germany}

\affiliation{Institute for Advanced Simulation, Institut f\"ur Kernphysik,
and
J\"ulich Center for Hadron Physics, Forschungszentrum J\"ulich,
D-52425 J\"ulich, Germany}

\affiliation{Tbilisi State University, 0186 Tbilisi, Georgia}


\begin{abstract}
We propose a new Monte Carlo method called the pinhole trace algorithm for {\it ab initio} calculations of the thermodynamics of nuclear systems. 
For typical simulations of interest, the computational speedup relative to conventional grand-canonical ensemble calculations can be as large as a factor of one thousand.  Using a leading-order effective interaction that reproduces the properties of many atomic nuclei and neutron matter to a few percent accuracy,
we determine the location of the critical point and the liquid-vapor coexistence line for symmetric nuclear matter with equal numbers of protons and neutrons.
We also present the first {\it ab initio} study of the density and temperature dependence of nuclear clustering.
\end{abstract}

\maketitle

In recent years much progress has been made in \textit{ab
initio} or fully microscopic calculations of the structure of atomic
nuclei \cite{Stroberg:2016ung,Piarulli:2017dwd,Lonardoni:2017hgs,
Gysbers:2019uyb,Smirnova:2019yiq,Contessi2017}.
These first principles calculations are based on chiral effective field
theory, whereby nuclear interactions are included term by term in order of
importance \cite{Epelbaum:2008ga}.
Unfortunately, most \textit{ab initio}
methods rely on computational strategies that are not designed for calculations
at nonzero temperature. 
One exception is many-body perturbation theory where
diagrammatic expansions are used to calculate bulk thermodynamic properties \cite{Baldo:1999cvh,Holt:2013fwa}.
Another exception is the method of self-consistent Green's functions, which provides non-perturbative solutions of the finite temperature system ~\cite{Soma:2009pf,Carbone:2018kji,Carbone:2019pkr,Carbone:2020}.
As with most first principles methods, however, these approaches have difficulties describing cluster correlations, which is an important feature of nuclear multifragmentation and the phase diagram of nuclear matter.

Yet another exception, which we focus on here, is the method of lattice effective field theory.
Lattice effective field theory has the advantage
that non-perturbative effects such as clustering are reproduced automatically when using Monte Carlo simulations.
Early efforts to describe nuclear thermodynamics using
lattice simulations exist in the literature~\cite{Muller:1999cp,Lee:2004si}, but there has been
little progress since then.  The difficulties stem from the computational
cost of performing grand-canonical calculations of nucleons in large
spatial volumes.  One can reduce the effort by working in a restricted
single-particle space~\cite{Alhassid:2014fca,Gilbreth:2019gkb}. Fully unbiased calculations, 
however, require a great amount of effort as they use matrices of size $4V \times 4V$, where
$V$ is the spatial lattice volume.  In this Letter, we report a new paradigm for calculating
\textit{ab initio} nuclear thermodynamics, which we call the pinhole trace algorithm.
In this algorithm, the matrices are much smaller, namely of size $A \times A$, where $A$ is 
the number of nucleons. The resulting computational acceleration can be as large as a factor of one thousand.

The \textit{ab initio} calculations presented here use the pinhole trace algorithm
to implement nuclear lattice effective field theory (NLEFT)~\cite{Lee2009, Lahde2019}
at finite temperature. 
At fixed nucleon number $A$, and temperature $T$, the expectation value of any observable $\mathcal{O}$
is
\begin{equation}
\langle\mathcal{O}\rangle_{\beta}=\frac{Z_{\mathcal{O}}(\beta)}{Z(\beta)}=\frac{{\rm Tr}_{A}(e^{-\beta H}\mathcal{O})}{{\rm Tr}_{A}(e^{-\beta H})},\label{eq:thermal_average}
\end{equation}
where $Z(\beta)$ is the canonical partition function, $\beta=T^{-1}$ is the inverse temperature, $H$ is the Hamiltonian, and ${\rm Tr}_{A}$ is the trace over the $A$-nucleon states.  Throughout, we use units where 
$\hbar = c = k_B = 1$.


The canonical partition function $Z(\beta)$ can be written explicitly in the single particle basis as
\begin{eqnarray}
   Z(\beta)& = &\sum_{c_1, \cdots , c_A} \langle c_1, \cdots , c_A | \exp(-\beta H)
     |  c_1, \cdots, c_A \rangle, \label{eq:partitionfunction}
\end{eqnarray}
where the basis states are Slater determinants composed of point particles, $c_i = ({\bm{n}_i, \sigma_i, \tau_i})$ are the quantum numbers of the $i$-th particle, with $\bm{n_i}$ an integer triplet specifying the lattice coordinate,
$\sigma_i$ is the spin and $\tau_i$ is the isospin.
On the lattice, the components of $\bm{n_i}$ take integer values from 0 to $L - 1$, where $L$ is the box length in units of the lattice spacing.
The neutron number $N$ and proton number $Z$ are separately conserved, and the summation in Eq.~(\ref{eq:partitionfunction}) is limited to the subspace with the specified values for $N$ and $Z$.

In the Supplemental Materials we present the full details of the lattice calculations.  But in order to explain the basic design of our computational approach, we illustrate here a simplified calculation where the Hamiltonian has a two-body contact interaction
\begin{equation}
    H = H_{\rm free} + \frac{1}{2} C \sum_{\bm n} : \rho^2({\bm n}) :,
\end{equation}
where $H_{\rm free}$ is the free nucleon Hamiltonian with nucleon mass $m=938.9$ MeV, $\rho({\bm n}) = \sum_{\sigma, \tau} \hat{a}^{\dagger}_{{\bm n}, \sigma, \tau} \hat{a}_{{\bm n}, \sigma, \tau}$ is the density operator.  The $::$ symbols indicate normal ordering where the annihilation operators are on the right and creation operators are on the left.
We assume an attractive interaction with $C < 0$. 

The imaginary time direction, whose length is set by the inverse temperature $\beta$, is divided into $L_t$ slices with time lattice spacing $a_t$ such that $\beta = L_t a_t$. For each time slice the two-body interaction is decomposed using an auxiliary-field transformation such that at each lattice site we have
\begin{equation}
   \exp\left (-\frac{a_tC}{2} \rho^2 \right) = \sqrt{\frac{1}{2\pi}} \int d s \exp \left( -\frac{s^2}{2} + \sqrt{-a_tC} s \rho \right),
\end{equation}
where $s$ is the auxiliary field. 

Putting these pieces together, we obtain the (auxiliary-field) path-integral expression for Eq.~(\ref{eq:partitionfunction})
\begin{eqnarray}
    Z(\beta) &=& \sum_{c_1, \cdots, c_A} \int \mathcal{D}s_{1} \cdots \mathcal{D} s_{L_t} \langle c_1, \cdots, c_A | \times \\
             & & M(s_{L_t}) \cdots M(s_1) |  c_1, \cdots, c_A \rangle, \label{eq:partitionfunction_pathintegral}
\end{eqnarray}
where
\begin{equation}
    M(s_{n_t}) = : \exp\left[-a_t K + \sqrt{-a_t C} \sum_{{\bm n}} s_{n_t}({\bm n}) \rho({\bm n})\right] :
    \label{eq:transfermatrix}
\end{equation}
is the normal-ordered transfer matrix for time step $n_t$, and $s_{n_t}$ is our shorthand for all auxiliary fields at that time step \cite{Lee2009,Lahde2019}.
$K = -\nabla^2 / 2M$ is the kinetic energy operator, which is discretized using finite difference formulae~\cite{Lee2009}.
For a given configuration $s_{n_t}$, the transfer matrix $M(s_{n_t})$ consists of a string of one-body operators
which are directly applied to each single-particle wave function in the Slater determinant.
For notational convenience, we will use the abbreviations $\vec{c} = \{c_1, \cdots, c_A\}$ and $\vec{s} = \{s_1, \cdots, s_{L_t}\}$.

The pinhole trace algorithm (PTA) was inspired by the pinhole algorithm used to sample the spatial positions and spin/isospin of the nucleons \cite{Elhatisari2017}.
However, the purpose, implementation, and underlying physics of the PTA for nuclear thermodynamics are vastly different from the original pinhole algorithm used for density distributions. In the PTA we evaluate Eq.~(\ref{eq:partitionfunction_pathintegral}) using Monte Carlo methods, i.e. importance sampling is used to generate an ensemble $\Omega$ of $\{\vec{s}, \vec{c}\}$ of configurations according to the relative probability distribution
\begin{equation}
    P (\vec{s}, \vec{c}) = \left| \braket{\vec{c}|  M(s_{L_t}) \cdots M(s_1) | \vec{c}} \right|.  \label{eq:PTAprob}
\end{equation}
The expectation value of any operator $\hat {O}$ can be expressed as
\begin{equation}
    \langle \hat{O} \rangle = \langle \mathcal{M}_O (\vec{s}, \vec{c}) \rangle_\Omega / \langle \mathcal{M}_1 (\vec{s}, \vec{c}) \rangle_\Omega,
    \label{eq:measurement}
\end{equation}
where
\begin{eqnarray}
   \mathcal{M}_O(\vec{s}, \vec{c}) &=& \langle\vec{c}| M(s_{L_t}) \cdots M(s_{L_t/2+1})\hat{O} \times \\
                                   & & M(s_{L_t/2}) \cdots M(s_1)| \vec{c} \rangle / P (\vec{s}, \vec{c}).
\end{eqnarray}

To generate the ensemble $\Omega$ we use the Metropolis algorithm to update $\vec{s}$ and $\vec{c}$ alternately.
We first fix the nucleon configuration $\vec{c}$ and update the auxiliary fields $\vec{s}$. Starting from the rightmost time slice $s_1$, we update $s_1, \cdots, s_{L_t}$ successively, as detailed in Supplemental Materials.

 After updating $\vec{s}$, we then update the nucleon configuration $\vec{c}$.
 To that end, we randomly choose a nucleon $i$ and move it to one of its neighboring sites
 \begin{equation}
   c_i = \{\bm{n}_i, \sigma_i, \tau_i\} \rightarrow c'_i = \{\bm{n}'_i, \sigma_i, \tau_i\},
  \end{equation}
  or flip its spin,
   \begin{equation}
   c_i = \{\bm{n}_i, \sigma_i, \tau_i\} \rightarrow c'_i =  \{\bm{n}_i, -\sigma_i, \tau_i\}.
  \end{equation}
 The corresponding new nucleon configuration $\vec{c}\,'$ is accepted if
 \begin{equation}
     P(\vec{s}, \vec{c}\,') / P(\vec{s}, \vec{c}) > r'  \label{eq:cUpdate_Metropolis}
 \end{equation}
 with $0 \leq r' < 1$ a random number.
 Because in the $\vec{c}$ update only one nucleon is moved or spin flipped at a time, the successive configurations are correlated.
 Only when all nucleons have been updated do we obtain statistically independent configurations.
For calculations described here, we found that about 16 $\vec{c}$ updates for every $\vec{s}$ update
  produced the optimal sampling efficiency.

At low temperatures the signal in Eq.~(\ref{eq:partitionfunction}) may be overwhelmed by stochastic noise due to the notorious sign problem, i.e. to the almost complete cancellation between positive and negative amplitudes.
In auxiliary-field simulations with attractive pairing interactions, the sign problem is held in check by pairing symmetries.
For the case of spin pairing, this means that for any nucleon with quantum numbers $({\bm n}, \sigma, \tau)$, we can find another nucleon with $({\bm n}, -\sigma, \tau)$.  As the transfer matrix in Eq.~(7) is spin-independent, the pairing symmetry is preserved irrespective of the auxiliary fields.  A similar pairing symmetry also holds for isospin pairing, with $\tau$ and $-\tau$.
These pairing symmetries produce single-nucleon amplitude matrices with eigenvalues that come in complex conjugate pairs, such that the corresponding matrix determinants remain positive.

In the PTA, the nucleon positions and indices are allowed to explore unpaired configurations and could spoil the protection from sign oscillations provided by pairing symmetries.
Indeed, this possibility is one reason why the method had not been considered earlier, and why grand-canonical calculations have instead been used for the thermodynamics of nuclear systems as well as ultracold atoms \cite{Drut2006, Drut2008}. Fortunately, we find that this issue is not realized here.
For all temperatures considered in this Letter, we find that the sign problem is rather mild, as the positive sign configurations have stronger amplitudes due to the attractive pairing interactions. However, the sign problem will eventually reemerge for temperatures very low compared to the Fermi energy.  For interactions without pairing symmetries, the sign problem will be far more severe and appear even in auxiliary-field Monte Carlo calculations without pinholes.

For the values of $A$, $V$, and $L_t$ of interest in this work, the computational scaling of the PTA is $A^2VL_t$, while that for the grand-canonical algorithm described in Ref.~\cite{Blankenbecler:1981jt} is $AV^2L_t$.  Details of the computational scaling analysis can be found in the Supplemental Materials.  The cost savings of the PTA is a factor of $V/A$, and the speed up factor associated with the PTA will be as large as one thousand, depending on the lattice spacing and particle density.



Next we discuss the measurement of the observables.
While the energies and density correlation functions can be directly measured by inserting the corresponding operators in the middle time step as in Eq.~(\ref{eq:measurement}), we still need to design efficient algorithms for computing intensive variables, e.g., chemical potential $\mu$ or pressure $p$. 
This contrasts with grand-canonical ensemble calculations where the chemical potential is given as an external constraint.

In classical thermodynamics simulations, the Widom insertion method (WIM) \cite{Widom1963} is used to determine the statistical mechanical properties \cite{Binder1997, Dullens2005}. 
In the WIM we freeze the motion of the molecules and insert a test particle to the system and measure the free-energy difference, from which the chemical potential can be determined.            
The advantage of the WIM is that we do not need the total free energy, which would require an evaluation of the partition function.
In the PTA we encounter a similar problem. 
The absolute free energy can only be inferred with an integration of the energy from absolute zero, which induces large uncertainties.
To solve this problem, we adapt the WIM to the quantum lattice simulations, with the test particles substituted by fermionic particles or holes in the system.



For every configuration $\vec{c}$ generated in the PTA, we calculate the expectation values associated with adding one  nucleon or removing on nucleon.  We define
\begin{eqnarray}
    \mathcal{B}_1 &=& \sum_{c'} \braket{ \vec{c}\cup c'| M(s_{n_t})\cdots M(s_{1}) | \vec{c}\cup c' } / P(\vec{s}, \vec{c}),   \nonumber \\
    \mathcal{B}_{-1} &=& \sum_i \braket{\vec{c} \setminus c_i | M(s_{n_t})\cdots M(s_{1}) | \vec{c}\setminus c_i}/ P(\vec{s}, \vec{c}), \label{eq:widomAmplitudes}
\end{eqnarray}
where the summation over $c'$ runs over all single particle quantum numbers and the summation over $i$ runs over all existing particles.
$P(\vec{s}, \vec{c})$ is the probability given in Eq.~(\ref{eq:PTAprob}).
The extra free energy of inserting or removing one particle is given by
\begin{equation}
    F(A \pm 1) - F(A) = -T \ln \left[\frac{\langle \mathcal{B}_{\pm 1} \rangle_\Omega}{ (A \pm 1)!} \right]   .
\end{equation}
Using the symmetric difference, we have
\begin{equation}
    \mu = [F(A+1) - F(A-1)] / 2 = \frac{T}{2} \ln \left[ A(A+1) \frac{ \langle \mathcal{B}_{-1} \rangle_\Omega } { \langle \mathcal{B}_{1} \rangle_\Omega } \right].
\end{equation}

In the PTA the summations in Eq.~(\ref{eq:widomAmplitudes}) can be calculated using random sampling.
For $\mathcal{B}_1$ we insert a nucleon with random spin and location and propagate it through all time slices, while for $\mathcal{B}_{-1}$ we simply remove one of the existing nucleon.
As only one particle is inserted/removed in each measurement, we find this algorithm very efficient and precise in calculating the chemical potential $\mu$.
Subsequently, we determine the pressure $p$ by integrating the Gibbs-Duhem equation, $dp = \rho d\mu$, starting from the vacuum with $p=0, \rho=0$.

\begin{figure}
\begin{centering}
\includegraphics[width=0.85\columnwidth]{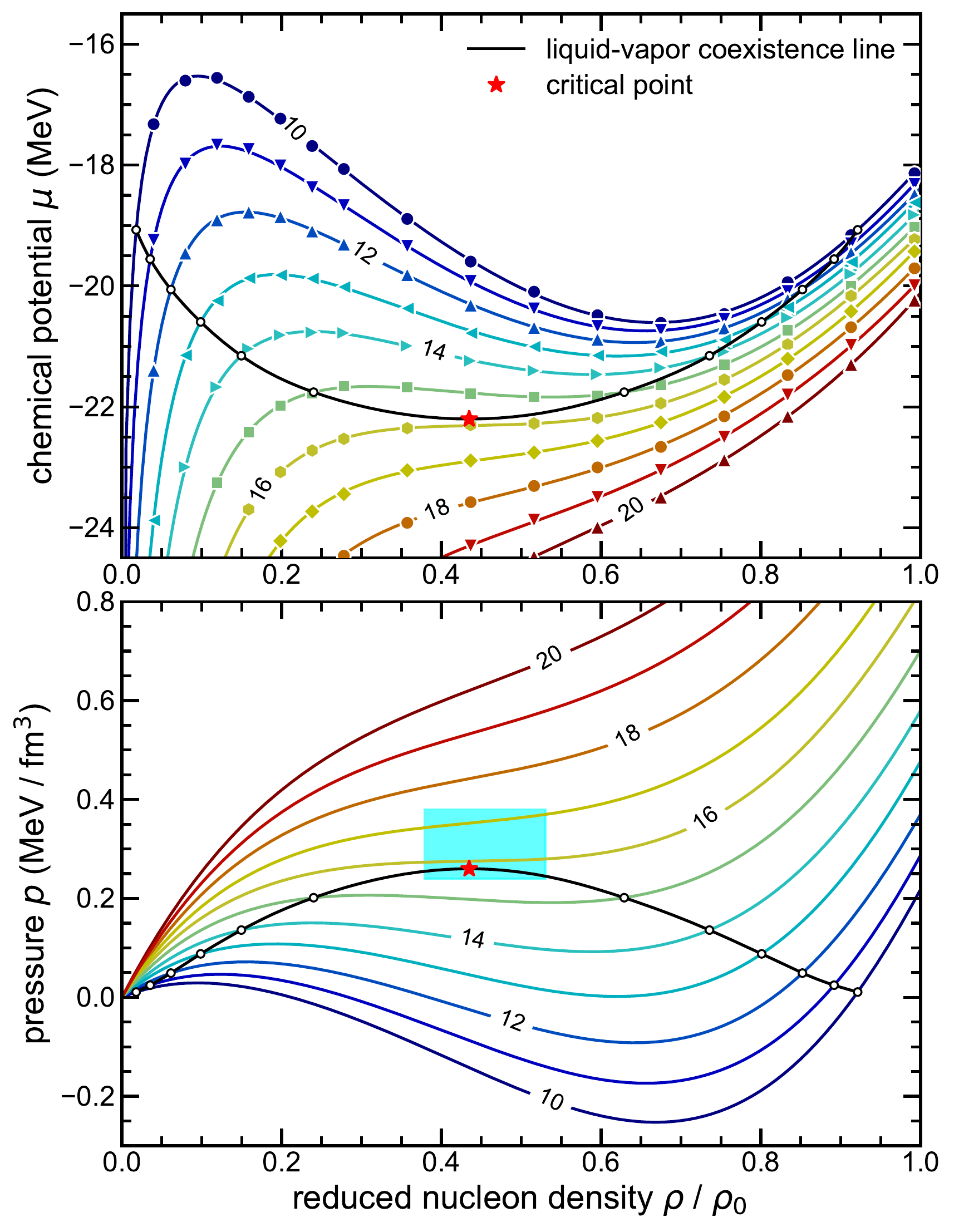}
\par\end{centering}
\caption{\label{fig:isotherms}
 (Upper panel) The $\mu$-$\rho$ isotherms of symmetric nuclear matter computed on the lattice
with $L^3=6^3$. The numbers on the lines are temperatures in MeV, and the temperature difference between adjacent isotherms is 1~MeV. The black
line denotes the liquid-vapor coexistence line
derived from Maxwell construction, and the red star marks the calculated
critical point.
    (Lower panel) The $p$-$\rho$ isotherms of symmetric nuclear matter are shown for $L^3=6^3$. The black
line denotes the liquid-vapor coexistence line, and the red star marks the calculated
critical point.
    The cyan rectangle marks the empirical critical point extracted from heavy-ion collisions \cite{Elliott2013}.
}
\end{figure}

As the long-range Coulomb interaction is ill-defined in the thermodynamic limit without screening, it is standard practice to remove the Coulomb force from nuclear matter calculations.  We note that in actual heavy-ion collisions the Coulomb interaction can be important, and so the comparison with Coulomb-removed nuclear matter is not entirely straightforward.
We first focus on the nuclear equation of state at nonzero temperatures, which is important for describing the evolution and dynamics of
core-collapse supernovae \cite{Togashi2017}, neutron star cooling \cite{Page2004},
neutron star mergers \cite{Most:2018eaw} and heavy-ion collisions \cite{Das:2004az}.  We then consider nuclear clustering as a function of density and temperature.


In this work we perform simulations on $L^3 = 4^3, 5^3, 6^3$ cubic lattice with up to 144 nucleons and a spatial lattice spacing $a=1/150$~MeV$^{-1}\approx1.32$~fm, such that the corresponding momentum cutoff is $\Lambda=\pi/a\approx471$~MeV. The temporal lattice spacing is taken to be $a_{t}=1/2000$~MeV$^{-1}$.
For these calculations we use the pionless effective field theory Hamiltonian introduced in Ref.~\cite{Lu:2018bat}, consisting of two- and three-body contact interactions which reproduce the binding energy and charge distribution of many light and medium-mass nuclei.  While this is a simple leading order theory and the results are only a first step towards higher-order calculations in chiral effective field theory, our simplified calculation has the important dual purpose of making our discussion of the PTA accessible to a broader audience for potential applications to the thermodynamics of condensed matter systems and ultracold atoms.

We impose twisted boundary conditions along the coordinate
directions, which means that each nucleon momentum component $p_i$ must equal $\theta_i/L+2\pi n_i/L$ for our chosen twist angle $\theta_i$ and some integer $n_i$.  As detailed in Supplemental Materials, we average each observable over all possible twist angles by Monte Carlo sampling.  As others have found \cite{Lin2001,Hagen:2013yba,Schuetrumpf:2016uuk}, twist averaging significantly accelerates the convergence to the thermodynamic limit.


In Fig.~\ref{fig:isotherms} we present the calculated chemical potential and pressure isotherms for $L^3 = 6^3$.
Each point represents a separate simulation.  The temperature $T$ covers the range from 10~MeV to 20~MeV and densities from 0.0080~fm$^{-3}$ to 0.20~fm$^{-3}$. The statistical errors are very small, less than 0.02~MeV for $\mu$ and less than 0.002~MeV/fm$^3$ for $p$.  These are too small to be clearly visible in Fig.~\ref{fig:isotherms} and are not shown. The liquid-vapor coexistence line is determined through the Maxwell construction
of each isotherm and depicted as a solid black line in Fig.~\ref{fig:isotherms}.
The liquid-vapor critical point is then located by solving the equations $d\mu/d\rho
= d^2\mu/d\rho^2 = 0$.
The same process applied to the data for $L=4^3$ and $L=5^3$ in order to estimate the error associated with extrapolation to the thermodynamic limit.

In Table~\ref{tab:critical} we present the calculated critical temperature $T_c$, density $\rho_c$ and critical pressure $P_c$.
The first error bar represents the combined uncertainty from statistics and extrapolation to the thermodynamic limit. 
The second error bar is the estimated systematic uncertainty associated with the contribution of omitted higher-order interactions.
For completeness we also present the saturation density $\rho_{\rm sat}$ at $T = 0$ MeV and the saturation energy per nucleon $E_{\rm sat} / A$.
We compare our results with the perturbative calculations using N$^3$LO chiral interactions~\cite{Wellenhofer2014} with two different momentum cutoffs. There appears to be a significant amount of dependence on the momentum cutoff, and the difference gives a rough estimate of the corresponding uncertainties. 
In the last column we present the empirical values deduced from the heavy-ion collision experiments~\cite{Elliott2013}.

We note that while the empirical $\rho_{\rm sat}$ extracted from heavy-ion collisions is about $25\%$ lower than the standard value of $0.17$~fm$^{-3}$ and our lattice $\rho_{\rm sat}$ is about $25\%$ higher than $0.17$~fm$^{-3}$, the ratios for $\rho_{c}/\rho_{\rm sat}$ for the two cases are in agreement with each other and also in agreement with the two N$^3$LO chiral results.  This is consistent with the general expectation that small systematic errors in the density can be reduced by computing ratios of densities.


\begin{figure}
\begin{centering}
\includegraphics[width=0.85\columnwidth]{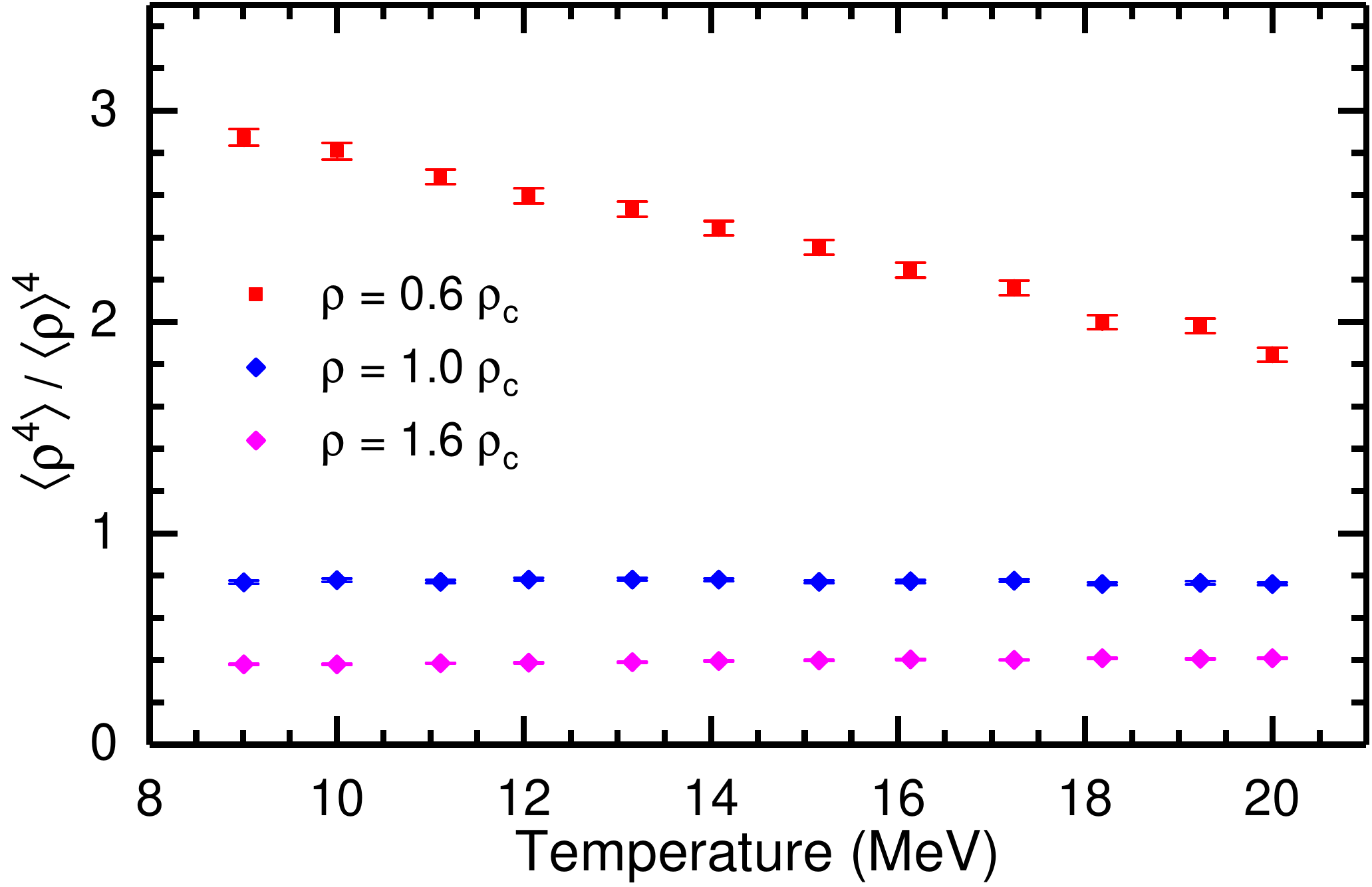}
\par\end{centering}
\caption{\label{fig:rho4}
The temperature dependence of the four-body density $\langle :\rho^4: \rangle$ scaled by the fourth power of the nucleon density $\langle\rho\rangle$.
The results are calculated for symmetric nuclear matter on the lattice with $L^3 = 5^3$.
The squares, circles, and diamonds represent total nucleon number $A = 16, 28, 40$, respectively.
The corresponding nucleon densities are $\rho = 0.6 \rho_c, 1.0 \rho_c, 1.6 \rho_c$, with $\rho_c = 0.089$ fm$^{-3}$ the calculated critical density.
    }
\end{figure}


\begin{table}
\caption{\label{tab:critical}
The calculated critical temperature $T_c$, pressure $P_c$, density $\rho_c$,
saturation density $\rho_{\rm sat}$ and energy per nucleon $E_{\rm sat} / A$.
For comparison we also present the results of perturbative calculations using N$^3$LO chiral potentials,
"n3lo414" and "n3lo500" correspond to cutoff momentum $\Lambda = 414$ and $500$ MeV, respectively~\cite{Wellenhofer2014}.
"Exp." denotes the empirical values inferred from the cluster distributions in the multi-fragmentation experiments~\cite{Elliott2013}.
}
\centering{}%
\begin{tabular}{cccccc}
\hline
 &  & This work & n3lo414 & n3lo500 & Exp.\tabularnewline
\cline{1-1} \cline{3-6}
$T_{c}$(MeV) &  & $15.80(0.32)(1.60)$ & $17.4$ & $19.1$ & $17.9(4)$\tabularnewline
$P_{c}$(MeV/fm$^{3}$) &  & $0.260(05)(30)$ & $0.33$ & $0.42$ & $0.31(7)$\tabularnewline
$\rho_{c}$(fm$^{-3}$) &  & $0.089(04)(18)$ & $0.066$ & $0.072$ & $0.06(1)$\tabularnewline
$\rho_{{\rm sat}}$(fm$^{-3}$) &  & $0.205(08)(40)$ & $0.171$ & $0.174$ & $0.132$\tabularnewline
$E_{{\rm sat}}/A$(MeV) &  & $-16.9(0.3)(1.7)$ & $-15.79$ & $-16.51$ & \tabularnewline
\hline
\end{tabular}
\end{table}

Nuclear clustering is another important phenomenon essential to our understanding of the phase diagram of nuclear matter and multifragmentation in heavy-ion collisons \cite{Ono:2019}.  Here we present the first study of nuclear clustering in a fully {\it ab initio} thermodynamics calculation.
Nuclear clustering is a manifestation of strong many-body correlations which goes well beyond mean-field theory, and thus is very difficult to reproduce using most {\it ab initio} methods.
In Fig.~\ref{fig:rho4} we show the expectation values of the four-body density $\langle :\rho^4: \rangle$ for different temperatures.
To build a dimensionless observable, we scale the results by the nucleon density $\langle \rho \rangle$ to the fourth power.
The resulting quantity $\kappa$ is a sensitive indicator of the degree of four-body clustering or alpha clustering.
Here we present the results for three different nucleon densities, which correspond to $0.6$, $1.0$ and $1.6$ times the critical density $\rho_c$.
For sub-critical density $\rho = 0.6\rho_c$ the system is a plasma of small clusters and we found $\kappa \gg 1$.
As the temperature increases the clusters begin to disintegrate and $\kappa$ decreases.
For the critical density $\rho = \rho_c$ and super-critical density $\rho = 1.6 \rho_c$, we found negligible alpha clustering with $\kappa < 1$.
Here the thermal motion and small interparticle spacing overwhelm the tendency for clustering in such hot and dense environments.
In this regime we find that alpha clustering is a monotonically decreasing function of the nucleon density, but does not depend on the temperature.
Since nuclear clustering is very difficult to probe using other {\it ab initio} methods, it would be extremely interesting and useful to build upon this first study and investigate the density and temperature dependence of nuclear clustering in more detail with the PTA and high-quality chiral nuclear forces.

Future work will improve upon these calculations by including higher-order interactions in lattice effective field theory. With the pinhole trace algorithm, many exciting applications are possible based on first principles calculations of quantum many-body systems at nonzero temperature.  This includes studies of superfluidity in symmetric and asymmetric nuclear matter, neutrino interactions in warm nuclear matter and supernova explosions, the properties of neutron stars and neutron star mergers, the temperature and density dependence of nuclear clusters, and extensions to other quantum many-body systems such as ultracold atoms and molecules.
\begin{acknowledgments}
We are grateful for discussions with Pawel Danielewicz, Christopher Gilbreth, and Bill Lynch. Through private discussions we have learned that Christopher Gilbreth is independently working on methods similar to the pinhole trace algorithm.  We acknowledge partial financial support from the Deutsche Forschungsgemeinschaft (TRR 110,
``Symmetries and the Emergence of Structure in QCD\textquotedblright ),
the BMBF (Verbundprojekt
05P18PCFP1), the U.S. Department of Energy (DE-SC0018638 and DE-AC52-06NA25396),
the National Science Foundation (grant no. PHY1452635), and the Scientific
and Technological Research Council of Turkey (TUBITAK project no.
116F400). Further support was provided by the Chinese Academy of Sciences
(CAS) President\textquoteright s International Fellowship Initiative
(PIFI) (grant no. 2018DM0034) and by VolkswagenStiftung (grant no.
93562). The computational resources were provided by the Julich Supercomputing
Centre at Forschungszentrum J\"ulich, Oak Ridge Leadership Computing
Facility, RWTH Aachen, and Michigan
State University.
\end{acknowledgments}

\beginsupplement
\onecolumngrid

\section{Supplemental Materials}

\subsection{Lattice Hamiltonian}

For calculating the nuclear equation of state and four-body clustering we
take the leading order pionless effective field theory presented in Ref.~\cite{Lu:2018bat}.
In this section we give the details.
On a periodic $L^3$ cube with lattice coordinates $\bm{n}=(n_{x,}n_{y},n_{z})$,
The Hamiltonian is
\begin{equation}
H_{{\rm SU(4)}}=H_{\rm free}+\frac{1}{2!}C_{2}\sum_{\bm{n}}:\tilde{\rho}^2(\bm{n}):+\frac{1}{3!}C_{3}\sum_{\bm{n}}:\tilde{\rho}^3(\bm{n}):,\label{eq:HSU4}
\end{equation}
where $H_{\rm free}$ is the free nucleon Hamiltonian with nucleon mass
$m=938.9$~MeV and the $::$ symbol indicate normal ordering.
 The density operator $\tilde{\rho}(\bm{n})$ is defined as
\begin{equation}
\tilde{\rho}(\bm{n})=\sum_{i}\tilde{a}_{i}^{\dagger}(\bm{n})\tilde{a}_{i}(\bm{n})+s_{L}\sum_{|\bm{n}^{\prime}-\bm{n}|=1}\sum_{i}\tilde{a}_{i}^{\dagger}(\bm{n}^{\prime})\tilde{a}_{i}(\bm{n}^{\prime}),
\end{equation}
where $i$ is the joint spin-isospin index and the smeared annihilation and
creation operators are defined as
\begin{equation}
\tilde{a}_{i}(\bm{n})=a_{i}(\bm{n})+s_{NL}\sum_{|\bm{n}^{\prime}-\bm{n}|=1}a_{i}(\bm{n}^{\prime}).
\end{equation}
The summation over the spin and isospin implies that the interaction is SU(4)
invariant. The parameter $s_L$ controls the strength of the local part of
the interaction, while $s_{NL}$ controls the strength of the nonlocal part
of
the interaction.
Here we include both kinds of smearing.  Both $s_L$ and $s_{NL}$ have an
impact on the range of the interactions.  The parameters $C_{2}$ and $C_{3}$
give the strength of the two-body
and three-body interactions, respectively.

In this letter we use a lattice spacing $a=1.32$~fm, which corresponds
to a momentum cutoff $\Lambda=\pi/a\approx471$~MeV. The dynamics
with momentum $Q$ much smaller than $\Lambda$ can be well described
and residual lattice artifacts are suppressed by powers of
$Q/\Lambda$. In this letter we use the parameter set $C_2 = -3.41\times 10^{-7}$
MeV$^{-2}$,
$C_3 = -1.4 \times 10^{-14}$ MeV$^{-5}$, $s_{L} = 0.061$ and $s_{NL} = 0.5$.
These parameters are adjusted to reproduce the deuteron, triton and the properties
of medium mass nuclei.

Nominally we are using a lattice action that can be viewed as a leading order
interaction in pionless effective field theory.  However it would be misleading
to associate the systematic errors of this work with that of other leading
order pionless studies such as Ref.~\cite{Contessi:2017rww}.  Our collaboration
has studied the importance of the range and strength of the local part of
the nucleon-nucleon interaction in order to reproduce the bulk properties
of nuclear matter \cite{Elhatisari:2016owd,Rokash:2016tqh}.  The interaction
used here utilizes this information in order to reproduce the binding energies
and radii of light and medium-mass nuclei and the equation of state of neutron
matter with no more than a few percent error.  All of the errors associated
with the systematic deficiencies of our interactions have been estimated
and presented in our results.
 

we calculate the NN S-wave phase shifts below relative
fit errors by comparing results with the
and

\subsection{Auxiliary field formalism}

We simulate the interactions of nucleons on the lattice using projection
Monte Carlo with auxiliary fields; see Ref.~\cite{Lee2009,Lahde2019} for
an overview
of methods used in lattice EFT. We use an auxiliary-field formalism
where the interactions among nucleons are replaced by interactions
of nucleons with auxiliary fields at every lattice point in space
and time. In the auxiliary-field formalism each nucleon evolves as
if it is a single particle in a fluctuating background of auxiliary
fields. We use a single auxiliary field at LO in the EFT expansion
coupled to the total nucleon density. The interactions are reproduced
by integrating over the auxiliary field. In our lattice simulations,
the spatial lattice spacing is taken to be $a$ = (150 MeV)$^{-1}$=
1.32 fm, and the time lattice spacing is $a_{t}$ = (2000 MeV)$^{-1}$=
0.0985 fm. For any fixed initial and final state, the amplitude for
a given configuration of auxiliary field is proportional to the determinant
of an $A\times A$ matrix $M_{ij}$. The entries of $M_{ij}$ are
the single nucleon amplitudes for a nucleon starting at state $j$
at $\tau$ = 0 and ending at state $i$ at $\tau$ = $\tau_{f}$.

We use a discrete auxiliary field that can simulate
the two-, three- and four-body forces simultaneously without sign oscillations.
This follows
from an exact operator identity connecting the exponential of the
two-particle density $\rho^{2}$ to a sum of the exponentials of the
one-particle density $\rho$:
\begin{equation}
:\exp\left(-\frac{1}{2}C\rho^{2}-\frac{1}{6}C_{3}\rho^{3}-\frac{1}{24}C_{4}\rho^{4}\right):=\sum_{k=1}^{N}\omega_{k}:\exp\left(\sqrt{-C}\phi_{k}\rho\right):\label{eq:Auxiliary_Field_Identity}
\end{equation}
where $C$ is the interaction coefficient, $C_{3}$ and $C_{4}$ are
coupling constants for three-body and four-body forces, respectively,
$\omega_{k}$'s and $\phi_{k}$'s are real numbers and the :: symbols indicate
the normal ordering of operators. In this
letter we only consider attractive two-body interactions with $C<0$.
In order to avoid the sign problem we further require $\omega_{k}>0$
for all $k$.  The implementation of higher-body interactions using auxiliary
fields is also discussed in Ref.~\cite{Korber:2017emn}.

To determine the constants $\phi_{k}$'s and $\omega_{k}$'s, we expand
Eq.~(\ref{eq:Auxiliary_Field_Identity}) up to $\mathcal{O}(\rho^{4})$
and compare both sides order by order. In the context of the nuclear
EFT, the three- and four-body interactions are usually much weaker
than the two-body interaction, and we use the following ansatz with $N=3$,
\begin{equation}
\omega_{1}=\frac{1}{\phi_{1}(\phi_{1}-\phi_{3})},\qquad\omega_{2}=1+\frac{1}{\phi_{1}\phi_{3}},\qquad\omega_{3}=\frac{1}{\phi_{3}(\phi_{3}-\phi_{1})}\label{eq:solutions}
\end{equation}
where $\phi_{2}=0$ and $\phi_{1}$ and $\phi_{3}$ are two
roots of the quadratic equation,
\begin{equation}
\phi^{2}+\frac{C_{3}}{\sqrt{-C^{3}}}\phi-\frac{C_{3}^{2}}{C^{3}}+(\frac{C_{4}}{C^{2}}-3)=0.\label{eq:N3
value-1}
\end{equation}
Using Vieta's formulas, it is straightforward to verify that Eq.~(\ref{eq:solutions})
satisfies Eq.~(\ref{eq:Auxiliary_Field_Identity}) up to $\mathcal{O}(\rho^{4})$.
For a pure two-body interaction $C_{3,4}=0$, the solution is simplified
to $\phi_{1}=-\phi_{3}=\sqrt{3}$, $\phi_{2}=0$, $\omega_{1}=\omega_{3}=1/6$,
$\omega_{2}=2/3$. The formalism Eq.~(\ref{eq:solutions}) is very
efficient in simulating the many-body forces non-perturbatively. The
corresponding auxiliary field $s(n_{t},\bm{n})$ only assumes three
different values $\phi_{1}$, $\phi_{2}$ and $\phi_{3}$ and can
be sampled with the shuttle algorithm described below.

\subsection{Updating the auxiliary field}

We update the auxiliary field $s(n_{t},\bm{n})$ using a shuttle algorithm,
in which only one time slice is updated at a time.
The shuttle algorithm works as follows.
(1) Choose one time slice $n_{t}$, record
the corresponding auxiliary field as $s_{{\rm old}}(n_{t},\bm{n}$).
(2) Update the corresponding auxiliary fields at each lattice site
$\bm{n}$ according to the probablity distribution $P\left[s_{{\rm new}}(n_{t},\bm{n})=\phi_{k}\right]=\omega_{k}$,
$k=1,2,3$. Note that $\omega_{1}+\omega_{2}+\omega_{3}=1$. (3) Calculate
the determinant of the $A \times A$ correlation matrix $M_{ij}$ using $s_{{\rm
old}}(n_{t},\bm{n})$
and $s_{{\rm new}}(n_{t},\bm{n})$, respectively. (4) Generate a random
number $r\in[0,1)$ and make the ``Metropolis test'': if
\[
\left|\frac{\det\left[M_{ij}\left(s_{{\rm new}}(n_{t},\bm{n})\right)\right]}{\det\left[M_{ij}\left(s_{{\rm
old}}(n_{t},\bm{n})\right)\right]}\right|>r,
\]
accept the new configuration $s_{{\rm new}}(n_{t},\bm{n})$ and update
the wave functions accordingly, otherwise keep $s_{{\rm old}}(n_{t},\bm{n})$.
(5) Proceed to the neighboring time slice, repeat steps 1)-4), and turn round
at the ends of the time
series.

The shuttle algorithm is well suited for small $a_{t}$. In this
case the number of time slices is large and the impact of a single
update is small. As the new configuation is close to the
old one, the acceptance rate is high.
We compared the
results with the HMC algorithm and found that the new algorithm is
more efficient. In most cases the number of independent configurations
per hour generated by the shuttle algorithm is usually three or four
times larger than that generated by the HMC algorithm.




\subsection{Computational scaling}
We discuss the computational scaling of our pinhole trace algorithm (PTA)
simulations and the comparison with grand canonical simulations based on
the well-known BSS method first described in Ref.~\cite{Blankenbecler:1981jt}.
 Both are determinant Monte Carlo algorithms for lattice simulations with
auxiliary fields, and so the comparison is relatively straightforward.  We
consider a system with $A$ nucleons, $V=L^3$ spatial lattice points, and
$L_t$ time steps.  We will drop constant factors such as the factor of 4
associated with the number of nucleon degrees of freedom. In the PTA we compute
$A$ single nucleon wave functions, where each wave function has $V$ components,
and the correlation matrix $M_{ij}$ will be an $A\times A$ matrix.  Meanwhile
the BSS algorithm requires computing $V$ single nucleon wave functions, where
each wave function has $V$ components, and the correlation matrix $M_{ij}$
will be a $V\times V$ matrix.  

For both algorithms, we update the auxiliary fields sequentially according
to time step.  We update all of the auxiliary fields at one time step before
moving on to the next time step. We now consider a full sweep that updates
all of the auxiliary fields.  During this sweep through the auxiliary fields,
the cost associated with updating the single nucleon wave functions is $AVL_t$
for the PTA and $V^2L_t$ for the BSS algorithm.  

For small $A$ we can update all of the $V$ auxiliary fields for a given time
step in parallel.  But as $A$ becomes large, we need to perform $A$ separate
updates per time step, with $V/A$ auxiliary fields updated at a time.  For
the PTA, the cost of calculating correlation matrices for the full update
over auxiliary fields is $A^2VL_t$, and the cost of calculating matrix determinants
for the full update is $A^4L_t$.  For the BSS algorithm, the cost of calculating
correlation matrices for the full update is $VL_t$, and the cost of calculating
matrix determinants for the full update is $AV^2L_t$.

The PTA has an additional update associated with the pinholes.  The cost
of calculating correlation matrices for the full update of all $A$ pinholes
is $A^2VL_t$, and the cost of calculating matrix determinants for the full
update is $A^4L_t$. 
For the values of $A$, $V$, and $L_t$ of interest in this work, the overall
computational scaling of the PTA is $A^2VL_t$, while that for the BSS algorithm
is $AV^2L_t$.  We see that the cost savings of the PTA is a factor of $V/A$.
 We find that the speed up associated with the PTA can be as large as one
thousand, depending on the lattice spacing and particle density.

\subsection{Autocorrelation and sign problem in the PTA}

\begin{figure}
\begin{centering}
\includegraphics[width=0.55\columnwidth]{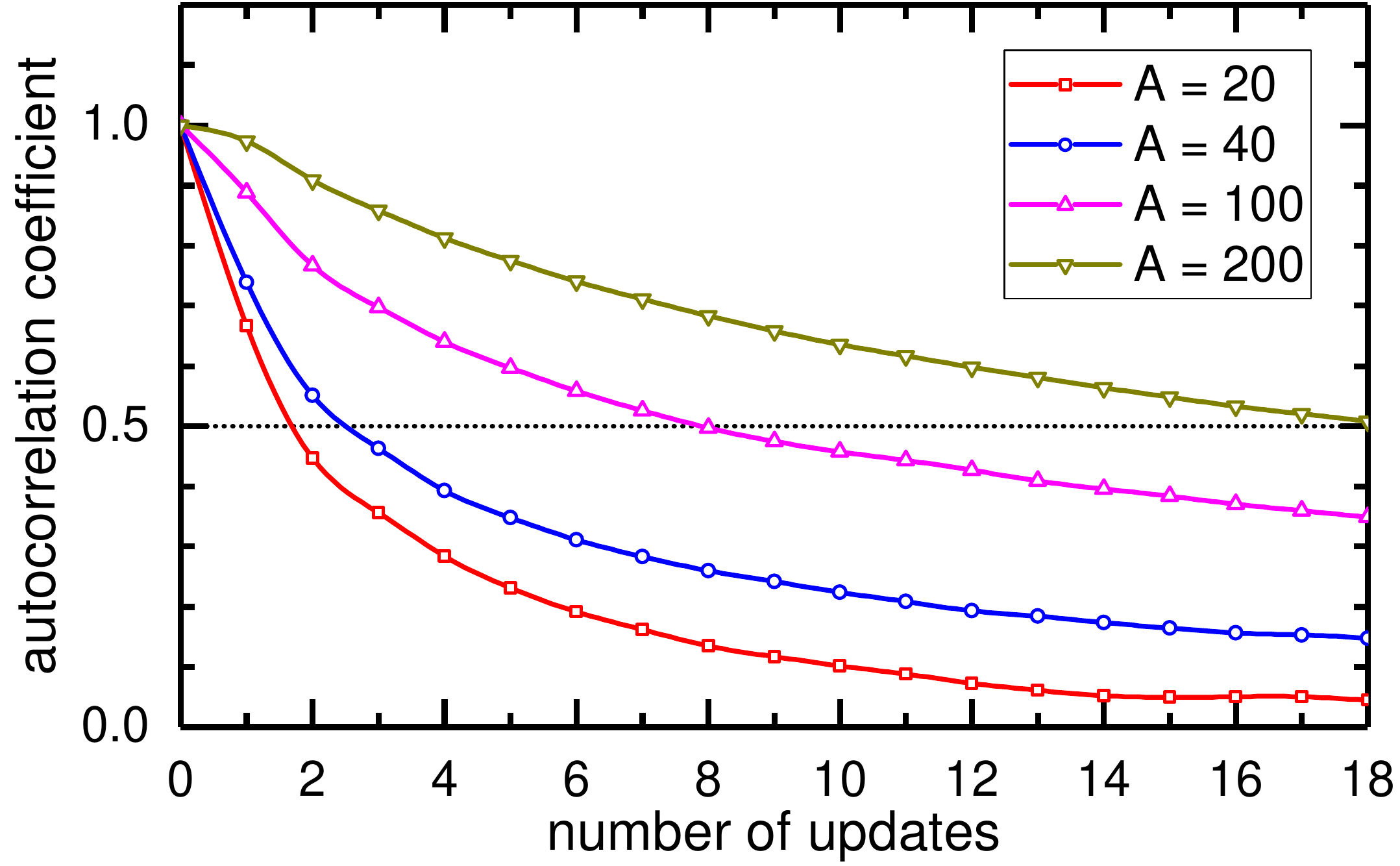}
\par\end{centering}
\caption{\label{fig:autocorrelation}
The auto-correlation coefficient of the total energy for the pinhole trace
algorithm.
$A$ is the number of nucleons, the temperature is $T=10$~MeV and box size
is $L=6$.
For each update cycle we update the auxiliary field once and pinhole configurations
16 times.
}
\end{figure}

In this section we test the performance of the PTA in terms of autocorrelation
and the sign problem.
 In Fig.~\ref{fig:autocorrelation} we show the auto-correlation coefficient
\begin{equation}
    \rho(n) = \frac{ {\rm Cov} [E(n_0), E(n_0 + n)] }
                   {  \sqrt{ {\rm D}[E(n_0)] {\rm D}[E(n_0 + n)] }  }   ,
\end{equation}
where $E(n_0)$ and $E(n_0 + n)$ are the energies calculated after $n_0$ and
$n_0 + n$ update cycles, respectively. ${\rm Cov}$ and ${\rm D}$ are covariance
and variance, respectively.
It is clearly seen that the auto-correlation time increases with the nucleon
number,
which means that for larger systems more $\vec{c}$ updates will be needed
to accelerate the convergence to equilibrium and reduce the uncertainties.
Nevertheless, in all calculations considered here, 16 $\vec{c}$ updates for
every $\vec{s}$ update
  appear to achieve optimal efficiency.

\begin{figure}
\begin{centering}
\includegraphics[width=0.55\columnwidth]{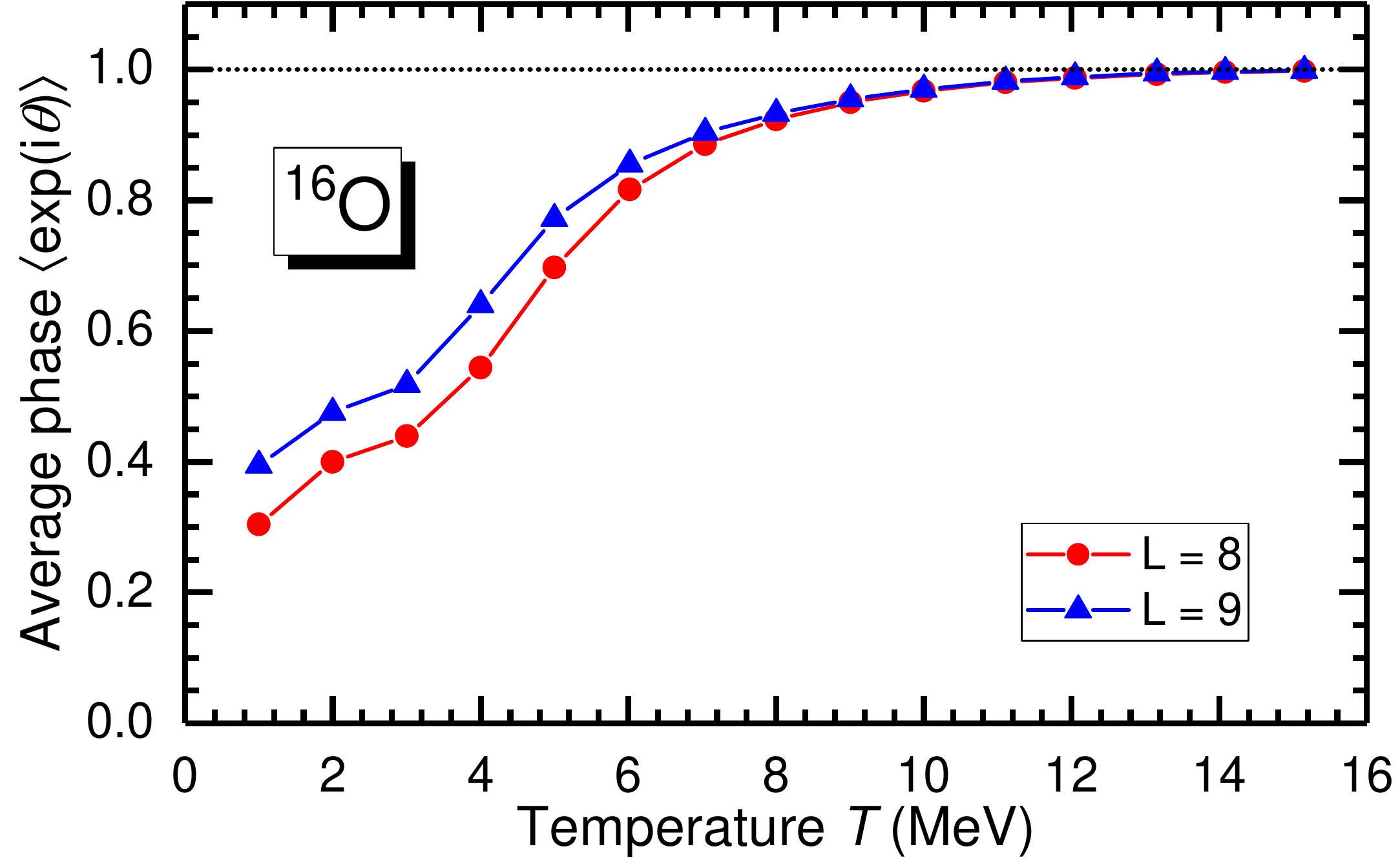}
\par\end{centering}
\caption{\label{fig:phase}
    The average phase $\langle e^{i \theta} \rangle$ as a function of the
temperature $T$ in $^{16}$O. Circles and triangles denote results for $L=8$
and $L=9$ respectively.
}
\end{figure}

At low temperatures the Monte Carlo simulations face the notorious sign problem,
which refers to cancellations between positive and negative amplitudes.
The average phase
\begin{equation}
  \langle e^{i \theta} \rangle = \langle \mathcal{M}_1 (\vec{s}, \vec{c})
\rangle_\Omega
\end{equation}
signifies the severity of the sign problem.
In practice the calculations become noisy when  $\langle e^{i \theta}\rangle$
is less than $0.1$.  In Fig.~\ref{fig:phase} we show the average phase of
PTA for temperatures from 1~MeV to 15~MeV in the $^{16}$O nucleus.
Here the average phase is a monotonically increasing function of the temperature
and asymptotes to 1 at high temperatures.
For temperatures as low as 1~MeV, the average phase decreases to $0.3$, which
requires a factor of ten times more measurements to achieve the same prescribed
precision.
Nevertheless, for all temperatures above 1~MeV, we find that the sign problem
is rather mild.

\begin{table}
    \caption{\label{tab:averagephase} Average phase $\langle e^{i\theta}\rangle$
in
an $L=6$ box. The Hamiltonian is the leading order pionless contact
    interactions in Ref.~\cite{Lu:2018bat}.}
\centering{}%
\begin{tabular}{ccccc}
\hline
$T$ (MeV) & $^{16}$O & $^{17}$O & $^{18}$O & $^{144}$Hf\tabularnewline
\hline
$10.0$ & 0.94 & 0.94 & 0.93 & 0.60\tabularnewline
$5.0$ & 0.47 & 0.45 & 0.43 & 0.02\tabularnewline
$2.0$ & 0.19 & 0.07 & 0.08 & 0.01\tabularnewline
\hline
\end{tabular}
\end{table}

In Table~\ref{tab:averagephase} we present the average phase $\langle e^{i\theta}
\rangle$ for several nuclei at different temperatures.
It is clear that the sign problem deteriorates for lower temperature or higher
density.
In all temperatures we find the largest phase for $^{16}$O.
The effect of broken pair in $^{17}$O and $^{18}$O is significant for $T=2.0$
MeV, but negligible for $T=5.0$ MeV and above.
This implies that the sign problem is due to the fermionic nature of the
nucleons, which is quenched either by pairing to form bosons or at high temperatures.

\subsection{Benchmark of the Widom insertion method}

\begin{figure}
\begin{centering}
\includegraphics[width=0.55\columnwidth]{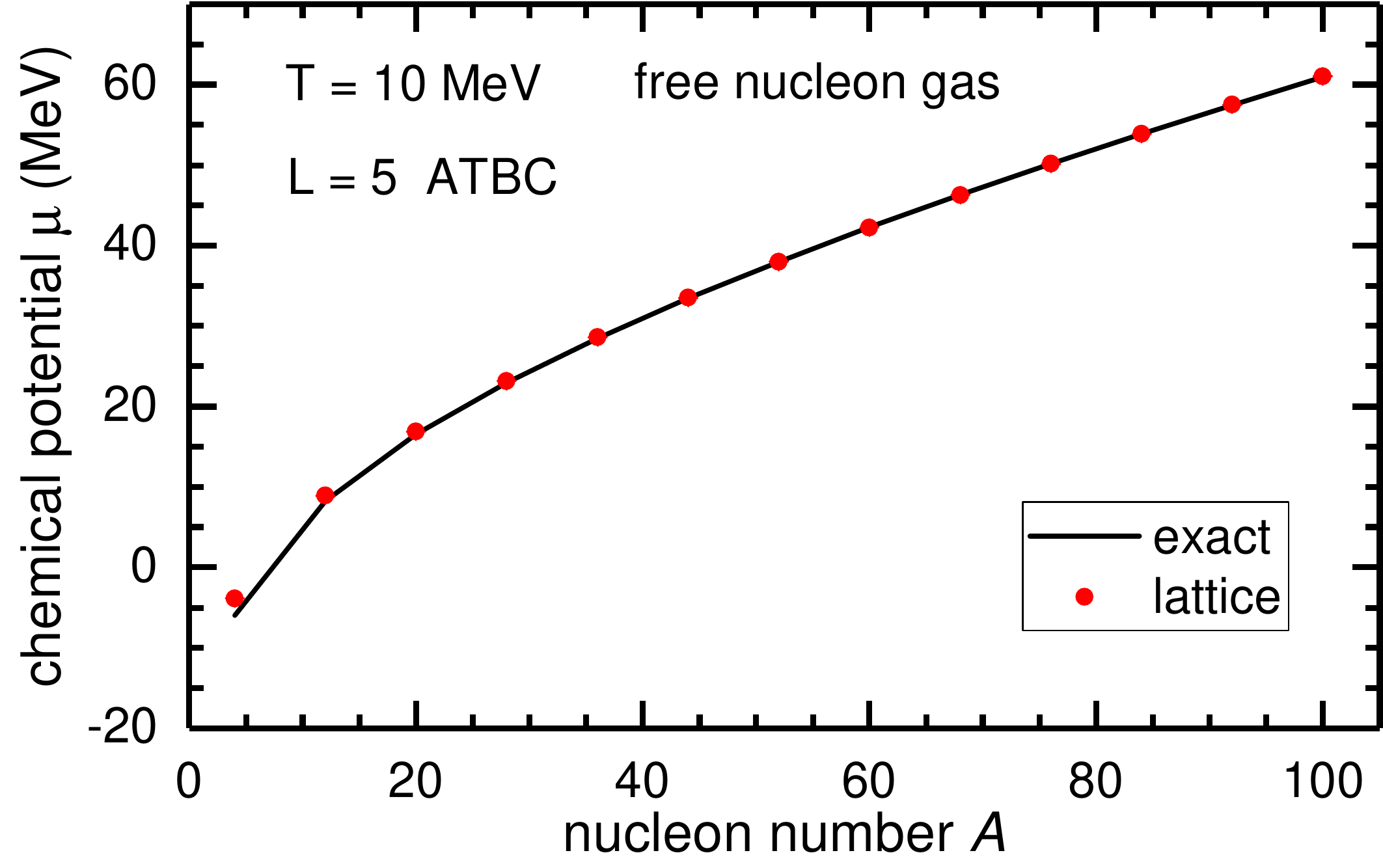}
\par\end{centering}
    \caption{\label{fig:ffg}
   The chemical potential of the free nucleon gas ($N$ = $Z$ = $A/2$) as
a function of the nucleon number $A$. The temperature $T=10$~MeV and the
box size $L=$5.
   Average twisted boundary conditions are applied in all directions (as
explained below). The circles denote the lattice results using the Widom
insertion method.
    The solid line shows the exact solution in the grand canonical ensemble.
}
\end{figure}

In this section we benchmark the WIM using a free nucleon gas where the chemical
potential can be calculated analytically.
In Fig.~\ref{fig:ffg} we compare the Monte Carlo results with the exact solutions.
In the grand canonical ensemble, the chemical potential $\mu$ can be determined
by solving the equation
\begin{equation}
    \int_0^\Lambda \frac{\rho(\epsilon)}{1 + e^{-\beta (\epsilon - \mu)}}
d\epsilon = A,
\end{equation}
where
\begin{equation}
    \rho(\epsilon) = \frac{2}{\pi^2} mV\sqrt{2m\epsilon}
\end{equation}
is the level density for free Fermi gas with two species, $m$ is the nucleon
mass, and $\Lambda = (\pi/a)^2/(2m)$ is the energy cutoff imposed by the
lattice at lattice spacing $a$.
In Fig.~\ref{fig:ffg} the solid line shows the exact solutions and the circles
show lattice results calculated using the QWIM.
The temperature is $T=10$~MeV, and the box size is $L=5$.
We see that the lattice results agrees well with the analytic solutions for
$A\leq 100$.
The deviation for small $A$ can be explained by the difference between the
canonical ensemble at fixed $A$ and the grand canonical ensemble at fixed
$\mu$.

\subsection{Backbending of the isotherms}

The calculated isotherms we have found follow the pattern expected for a
liquid-vapor phase transition in a finite system. Above $T_c$ the system
is in a supercritical state, while below $T_c$ the pure liquid and vapor
phases exist in the high- and low-density regime, respectively.
For states encompassed by the two arms of the coexistence line, the system
is a mixture of the liquid and vapor phases.
In the thermodynamic limit, where $N, V \rightarrow \infty$ with $\rho =
N/V$ fixed, $\mu$ and $p$ are constants in the coexistence regime along an
isotherm.
Both $\mu$ and $p$ are uniquely determined by the chemical and mechanical
equilibrium
conditions, $\mu_l = \mu_v = \mu_{\rm coex}$ and $p_l = p_v = p_{\rm coex}$,
where the subscripts $l$ and $v$ denote the liquid and vapor phases, respectively.
For a finite system the above conditions still hold. However, the surface
effects are non-negligible, and $\mu_{\rm coex}$ and $p_{\rm coex}$
can have different values.
A well-known example is that the pressure of the vapor in equilibrium with
small liquid drops can be larger than its thermodynamic-limit value,
with the difference compensated by the contribution of the surface tension.
Bearing the importance of the surface contributions in mind, we can easily
interpret the \textit{ab initio} calculations presented in Fig.~1
of the main text.

The most prominent feature of the isotherms in Fig.~1 is
the backbending curvature in the coexistence regime below $T_c$.
Note that the origin of this backbending is completely different from that
of similar structures found in the van der Waals model or other mean field
calculations.
The mean field models describe homogeneous systems and the backbending
of the $p$-$\rho$ isotherms result in a negative compressibility.  In this
regard the assumption of homogeneity conflicts with the condition
of mechanical equilibrium.
Conversely, in our \textit{ab initio} calculations we do not rely on the
assumption
of homogeneity; the results always describe realistic systems.
In particular, phase separation occurs spontaneously whenever it is favored
by the free energy criterion.
In the coexistence regime, the most probable configurations consist of high-density
liquid regions and low-density vapor regions, the surface separating
these regions
gives rise to a positive contribution to the total free energy, which prohibits
the formation of small liquid drop in vapor or small bubbles in liquid.
The distortions of the isotherms reflect the efforts of the system to overcome
such a surface energy barrier.

\begin{figure}
      \begin{centering}
        \includegraphics[width=0.55\columnwidth]{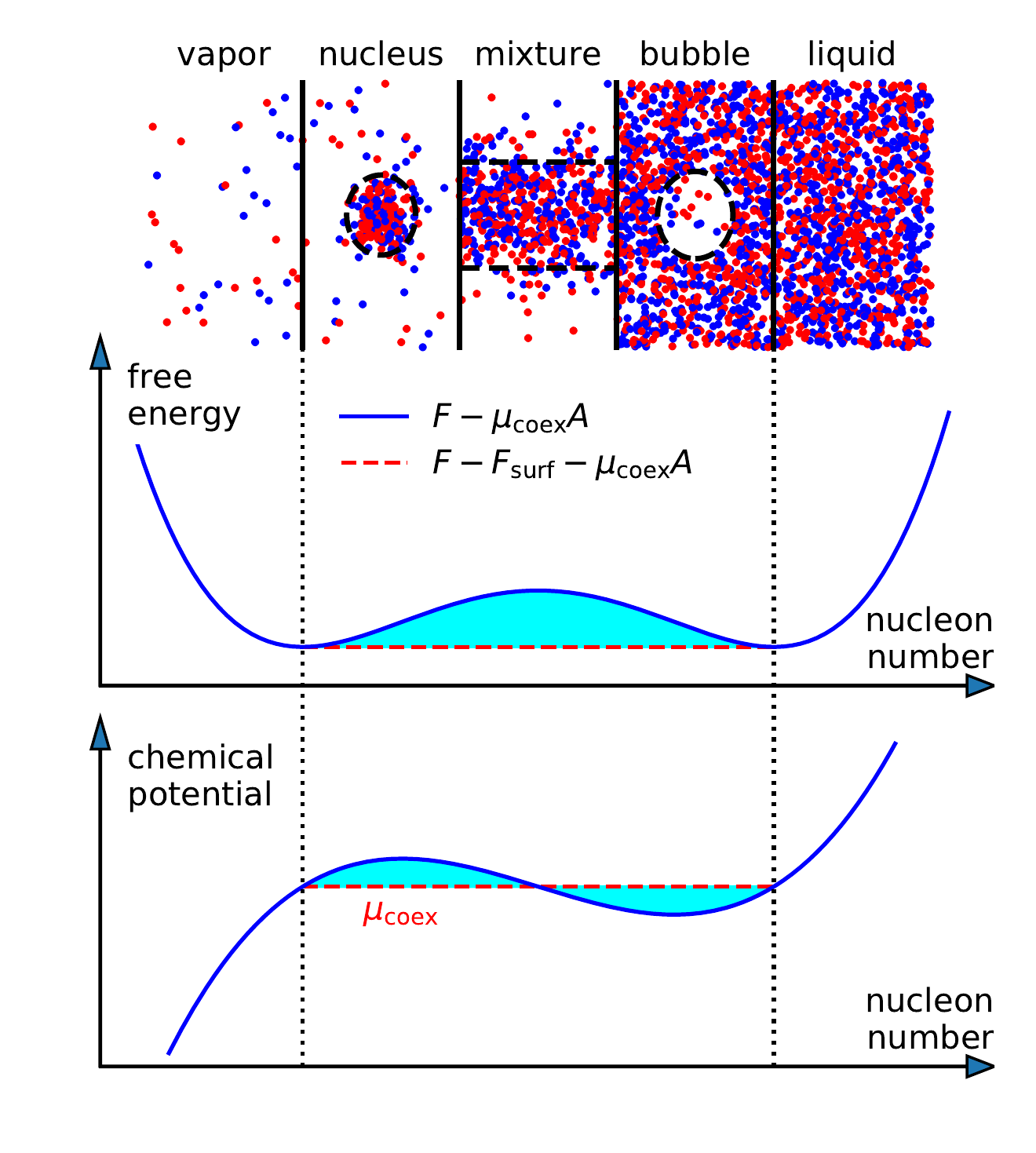}
        \par\end{centering}
        \caption{\label{fig:schematic}
          The schematic plots of the vapor-liquid phase transition in a finite
          nuclear system with fixed volume $V$ and temperature $T<T_c$.
          (Upper panel) The most probable configurations for different nucleon
          number $A$.
          Red (blue) points stand for protons (neutrons).
          The dashed lines signify the surfaces separating the liquid and
vapor
          phases.
          (Middle panel) The free energy.
          Solid (dashed) lines denote the results with (without) the surface
contributions.
          (Lower panel) The chemical potential $\mu = \partial{F}/\partial{A}$.
        }
    \end{figure}

In Fig.~\ref{fig:schematic} we show schematic plots illustrating the underlying
mechanism.
Given a fixed volume $V$ and a temperature $T$ below the critical value $T_c$,
the free energy $F$ is a function of the nucleon number $A$.
In the middle panel of Fig.~\ref{fig:schematic} we show the free energy curve
across the liquid-vapor coexistence region.
We subtract $\mu_{\rm coex} A$ from $F$ to remove most of the $A$-dependence,
with $\mu_{\rm coex}$ assuming the value at the thermodynamic limit.
For a finite system the surface free energy $F_{\rm surf}$ is approximately
proportional to the area of the surface.
In the upper panel of Fig.~\ref{fig:schematic} we show the most probable
configurations for different densities.
At low densities we have a nucleus surrounded by small clusters, while at
high densities we see bubbles in a nuclear liquid.
At intermediate densities the system contains bulk nuclear matter with appreciable
surface areas.
The surface area first increases after the formation of a nucleus then decreases
when most of the volume is occupied by the liquid phase.
Correspondingly $F_{\rm surf}$ has a unique maximum and creates a bump in
the free energy curve.
In the lower panel of Fig.~\ref{fig:schematic} we show the corresponding
chemical potential $\mu = \partial{F}/\partial{A}$.
The backbending is a natural result of the surface free energy
contributions.

\subsection{Finite volume effects}
In any first principles calculation of a quantum many-body system, we are
necessarily working with  finite number of nucleons in a finite volume. 
The finite volume together with the chosen boundary condition will induce
fictitious shell effects.
New lattice magic numbers for protons or neutrons emerge where the calculated
observables exhibit unphysical kinks.
It was observed that 66 particles for one species of spin-1/2 fermions give
results close to the thermodynamic limit.
This number was extensively used in most of the nuclear matter, neutron matter
or cold atom simulations \cite{Forbes2011,Carlson2011}.
However one would ideally like to explore different densities by varying
the number of nucleons as well as different the lattice volumes.
For this we must reduce as much as possible the problem of ficitious shell
effects.

The origin of the finite volume shell effects is the constraint imposed by
the boundary conditions on the particle momenta.
For a cubic box with periodic boundary conditions (PBC), particles are only
allowed to have momenta $\bm{p} = \frac{2\pi}{L}\bm{n}$,
which results in a series of magic numbers 2, 14, 38, $\cdots$ for one species
of spin-1/2 fermions.
One solution is to use twisted boundary conditions (TBC) \cite{Byers1961}
which attach extra phases to wave functions when particles cross the boundaries.
In this case the particle momenta are $\bm{p} = \bm{\theta} + \frac{2\pi}{L}\bm{n}$
for some chosen twist angles $\bm{\theta}$.  It has been found that averaging
over all possible twist angles provides an efficient way of approaching the
infinite volume limit \cite{Loh1988,Valenti1991,Hagen:2013yba,Schuetrumpf:2016uuk}.

The TBC method was first proposed for exactly solvable models \cite{Loh1988,
Valenti1991, Gros1992, Gammel1993, Gros1996} and then found applications
in quantum Monte Carlo methods \cite{Lin2001}.
Many groups have applied TBC to lattice QCD calculations to find infinite
volume results otherwise not accessible \cite{Bedaque2004, Divitiis2004,
Bedaque2005, Sachrajda2004}.
Meanwhile, the application of TBC to lattice effective field theory was shown
to be successful, though still limited to few-body and exactly solvable systems
\cite{Korber2016}. In this section we discuss the application of the TBC
to lattice Monte Carlo calculations and show how it helps remove finite-volume
shell effects in thermodynamics calculations.

  We apply the twisted boundary conditions to the single particle wave functions,
  \begin{eqnarray}
      \psi (x + L, y, z, \sigma, \tau) &=& \exp(i 2\sigma \theta_x) \psi
(x, y, z, \sigma, \tau), \nonumber \\
      \psi (x, y + L, z, \sigma, \tau) &=& \exp(i 2\sigma \theta_y) \psi
(x, y, z, \sigma, \tau), \nonumber \\
      \psi (x, y, z + L, \sigma, \tau) &=& \exp(i 2\sigma \theta_z) \psi
(x, y, z, \sigma, \tau),
  \end{eqnarray}
where $-\pi \leq \theta_x$, $\theta_y$, $\theta_z < \pi$ are the independent
twist angles in the three directions.
Note that for spin $\sigma = \pm 1/2$ we use the opposite twist angles, which
is necessary to preserve time reversal symmetry and avoid sign cancellations.
In this paper we employ TBC by averaging over all possible $(\theta_x, \theta_y,
\theta_z),$ which we call average twisted boundary conditions, ATBC.
This can be easily implemented in Monte Carlo calculations by allocating
to every thread a random phase triplet with elements uniformly distributed
in the interval $[-\pi, \pi)$.

\begin{figure}
\begin{centering}
\includegraphics[width=0.55\columnwidth]{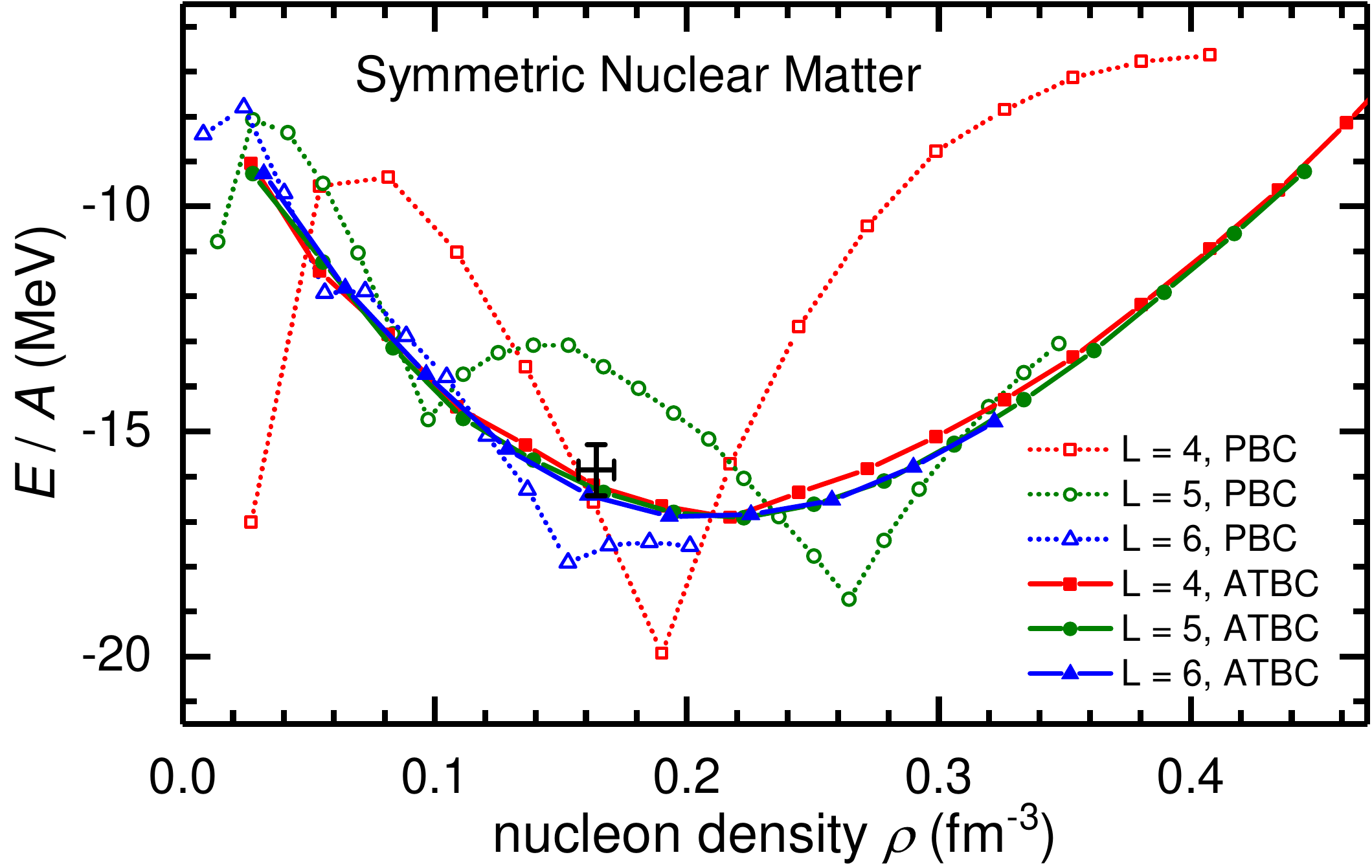}
\includegraphics[width=0.55\columnwidth]{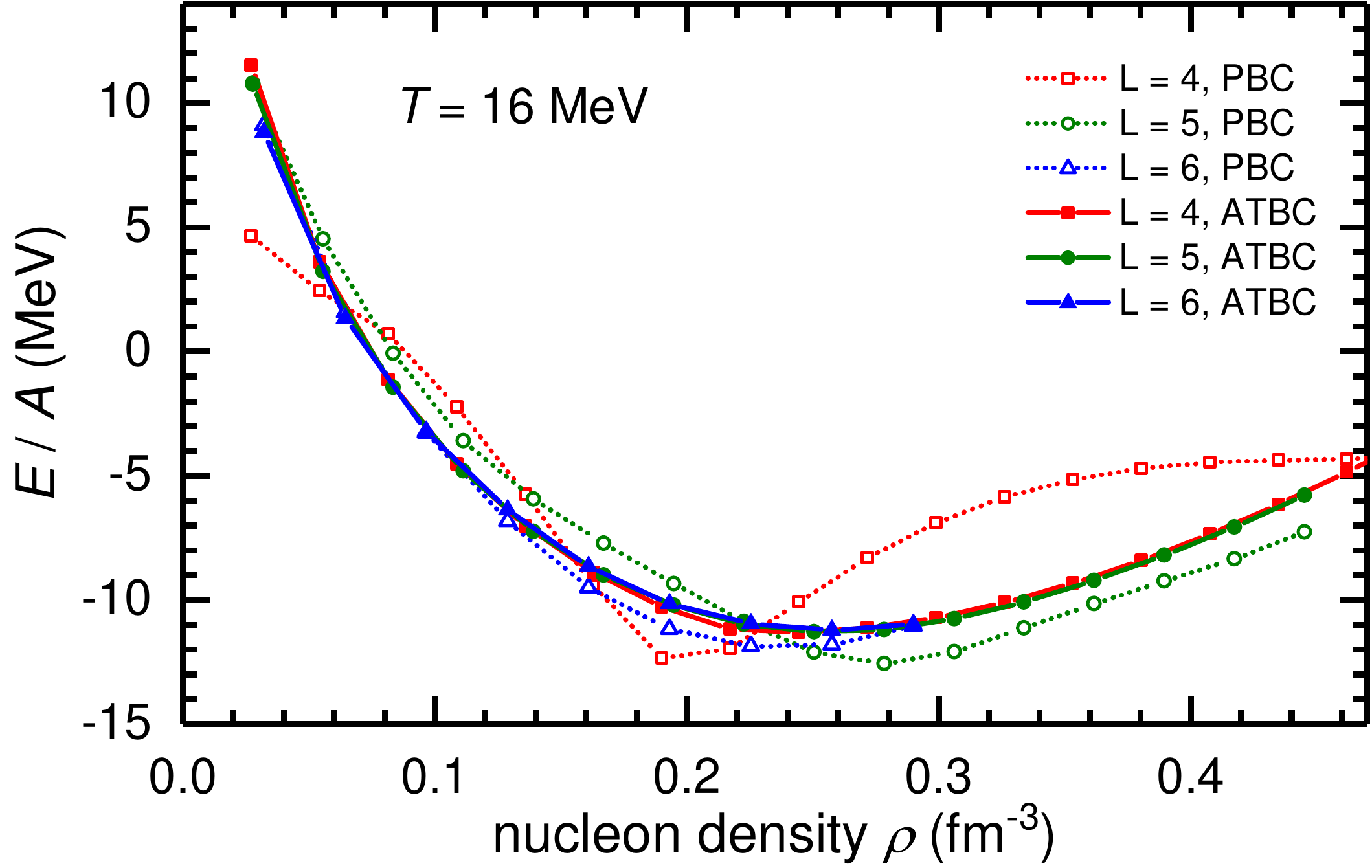}
\includegraphics[width=0.55\columnwidth]{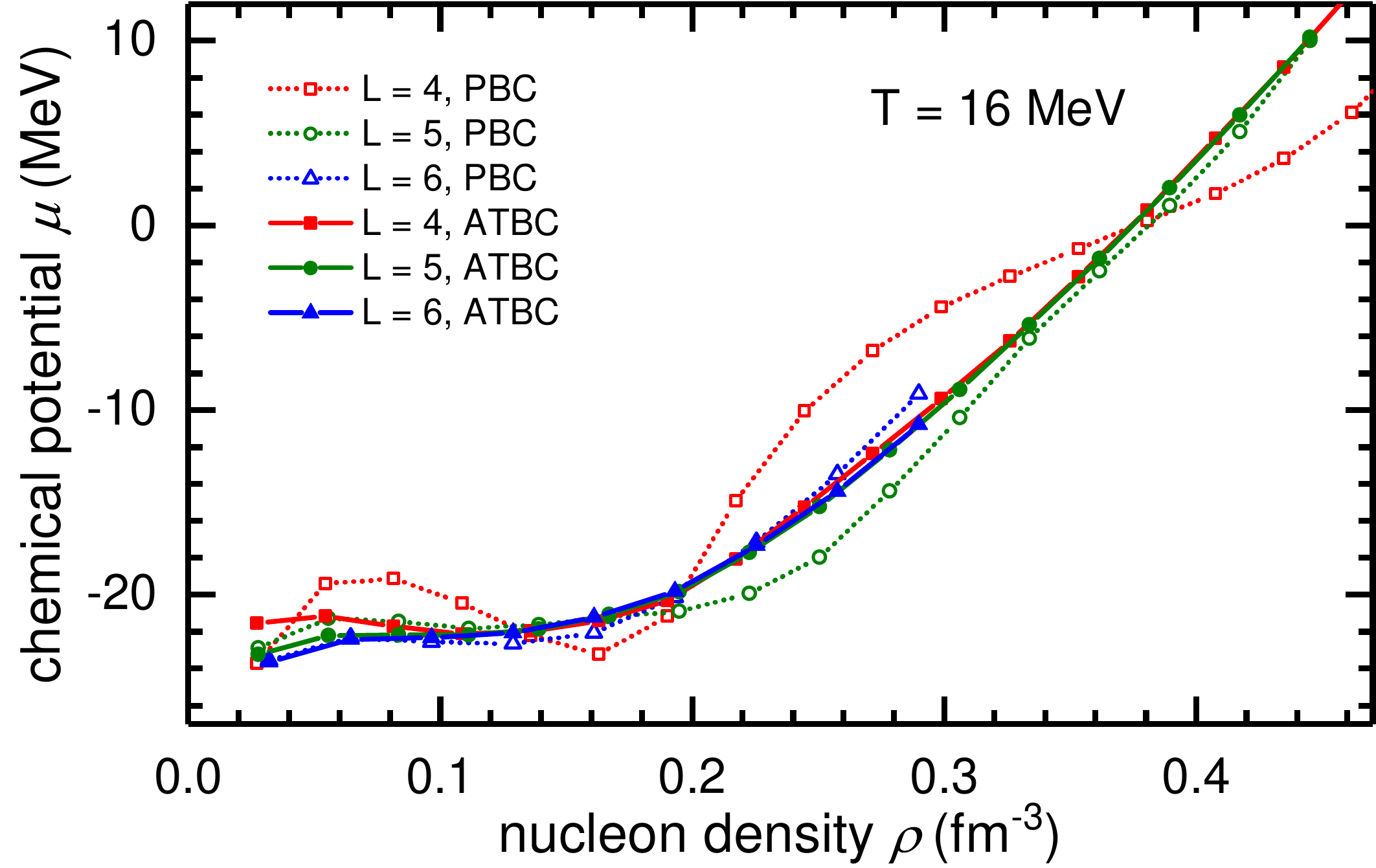}
\par\end{centering}
\caption{\label{fig:ATBC}
(Upper panel) The binding energy per nucleon calculated with periodic boundary
conditions (open symbols) and average twisted boundary conditions (full symbols,
as explained below).
    The temperature is $T=0$~MeV.
    The squares, circles and triangles show the results calculated with box
sizes $L=$ 4, 5 and 6, respectively.
    The cross with error bars shows the empirical saturation density and
binding energy.
    (Middle panel) The binding energy per nucleon for $T=$ 16~MeV.
    (Lower panel) The chemical potential at $T=16$~MeV.}
\end{figure}

In Fig.~\ref{fig:ATBC} we compare the binding energies at $T=0$ and $T=16$~MeV
and the chemical potential at $T=16$~MeV calculated with different boundary
conditions.
For the same density, different box sizes correspond to different nucleon
numbers.
The open symbols denote the results calculated with periodic boundary conditions,
the full symbols show the results for the average twisted boundary conditions.
Here we see clear shell effects for the PBC calculations.
For each box size the energy and chemical potential oscillates with respect
to the nucleon number and exhibit extrema at lattice magic numbers $A=$ 4,
28, 76, $\cdots$.
The amplitudes of the oscillation are smaller for larger boxes, but for $L=6$
the ficitious shell effects are still apparent.
For example, for $T=0$~MeV the energy minimum occurs at $\rho \approx 0.153$~fm$^{-3}$,
which is a shell effect that corresponds to $A=76$.
These results can be misleading if we do not take into account the finite
volume corrections.

With ATBC, each of the kinks found above disappear and the results collapse
onto universal curves. The chemical potential or the Fermi level, which is
very sensitive to the shell structure, can now be precisely calculated and
the results for $L=4,5,6$ can be used to estimate the uncertainty in the
extrapolation to the thermodynamic limit.

Some remarks must be added for the finite volume effects.
Here we distinguish between finite volume effects and finite size effects.
The former comes into play together with the boundary conditions and can
be removed by using twisted boundary conditions.
However, the latter is due to the finite particle number and manifests itself
mainly through the surface effects.
That is, the finite size effects are maximized for inhomogeneous systems,
in particular, the system comprising of two or more phases.
The contact surfaces of the different phases give positive contributions
to the free energy, which will vanish at the thermodynamic limit.
For example, for symmetric nuclear matter with sub-saturation density, the
system we described can be viewed as a large volume of liquid containing
a number of small bubbles, with bubble density $\rho=1/V$, where $V$ is the
volume used in the simulation.
For large $V$ at fixed density, the bubbles merge together into large ones
and the surface effects will eventually disappear.

The finite size effects scale with the surface-volume ratio, which in turn
scales as $O(A^{-1/3})$ with respect to the nucleon number.
Thus these effects decay very slowly and cannot be removed with present computational
settings.
One example of the surface effects are the upbending of the $T=0$ energy
curves at low densities in Fig.~\ref{fig:ATBC}.
For infinite nuclear matter, at the sub-saturation densities the density
and the binding energy per nucleon will be exactly the value at the saturation
point.
However, for finite systems the extra surface energy causes the upbending
of the energy curve and makes it converging to the binding energy per nucleon
of small nucleus in the vacuum, $E/A \approx -8$~MeV.

\begin{figure}
\begin{centering}
\includegraphics[width=0.55\columnwidth]{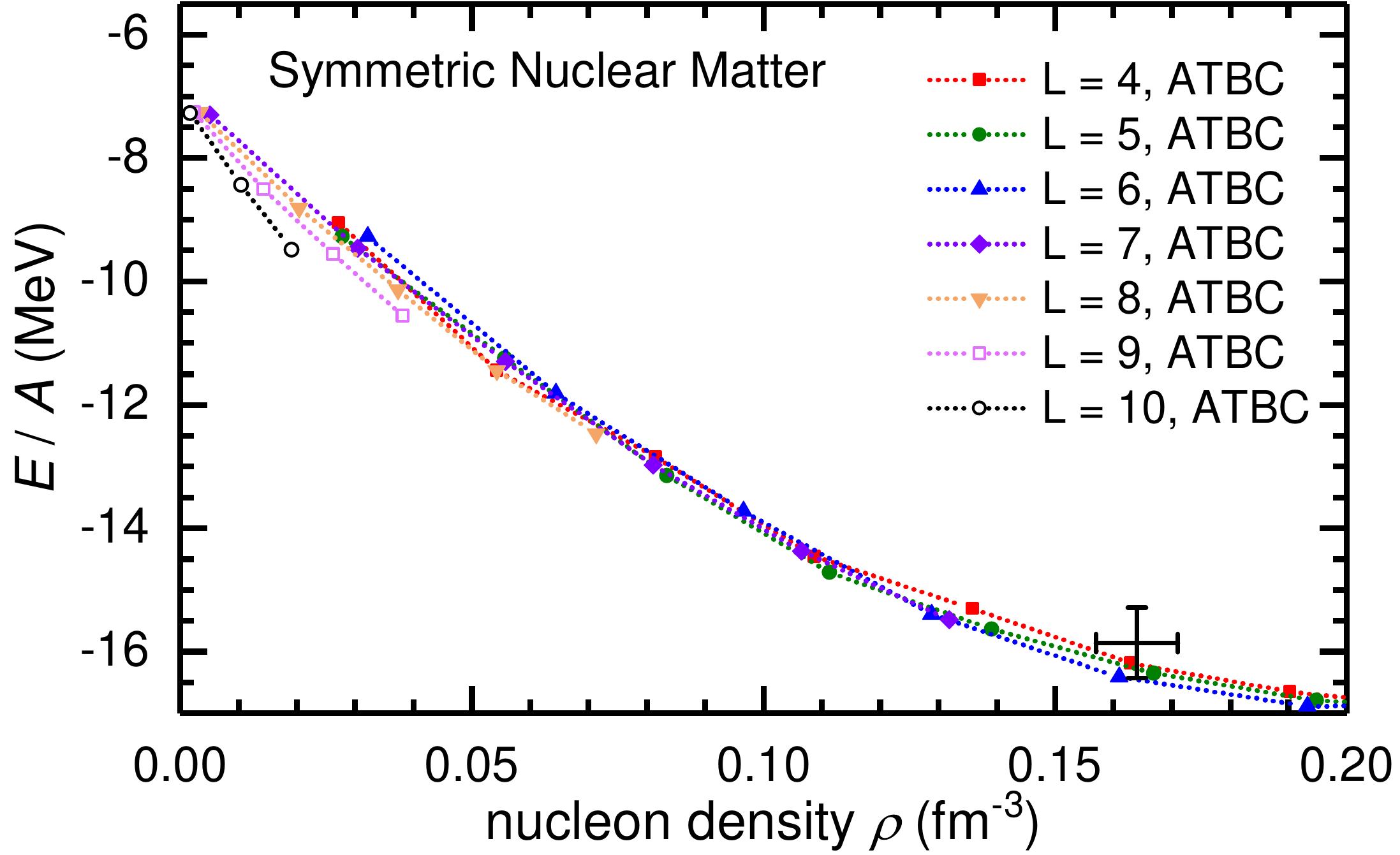}
\par\end{centering}
\caption{\label{fig:FSE}
    The energy per nucleon at $T=0$~MeV calculated with box sizes from $L=4$
to $L=10$.
    Average twisted boundary conditions are imposed in all directions.
    The cross with error bars shows the empirical saturation point.
   }
\end{figure}

In Fig.~\ref{fig:FSE} we show the energy per nucleon at $T=0$ calculated
with different box sizes.
In order to show the finite size effects we have removed the finite volume
effects by imposing the average twisted boundary conditions.
The calculated energies at sub-saturation energy tend to increase with respect
to the box size.
However, the convergence is extremely slow.
To reduce the finite size effects by one order of magnitude, we have to use
$10^3$ times more nucleons and $10^3$ times larger volume, which is not affordable
at present.

We must stress that the existence of the surface effects at phase coexistence
is not a deficiency of our method.
Instead of studying the infinite homogeneous matter, our method focuses on
the real finite systems with phenomena like cluster and phase separation.
Consequently, we believe that our formalism, together with the advanced nuclear
interactions, will pave the way of fully understanding the nuclear thermodynamic
processes.

\subsection{Interpolation and error analysis}
In extracting the equation of state and critical point, we make an interpolation
for the lattice results using the fifth-order virial expansion,
\begin{equation}
    \mu(\rho, T) = a_0 + a_1\ln \rho + \sum_{i=1}^{4} a_{i+1} \rho^i,
\end{equation}
where $a_i$ are functions of $T$ and should be fitted for each isotherm separately.
Note that this expression is only meant to parametrize the isotherms and
the resulting parameters cannot be compared directly with the real virial
coeffcients.
For non-integer temperatures the chemical potential is obtained by cubic
spline interpolation.
We the use the interpolated
equation of state for differentiation and integration.

The errors in the critical values are estimated by the bootstrap method whereby
we resample each lattice result with variance given by the Monte Carlo simulation
results several times, and estimate the variance of the critical values by
the resulting distribution.

\end{document}